\documentclass[article,superscriptaddress,aps,twocolumn,prx,quantum,longbibliography]{revtex4-2}
\usepackage{dcolumn}
\usepackage{bm}
\usepackage{epsfig}
\usepackage{graphicx}
\usepackage{amssymb,amsmath,amsbsy,amsgen,amsfonts,amsthm,mathtools}
\usepackage{mathrsfs}
\usepackage{latexsym}
\usepackage{array}
\usepackage{color}
\usepackage{xcolor}
\usepackage{amstext}
\allowdisplaybreaks[1]
\usepackage{txfonts}
\usepackage{pifont}
\usepackage{verbatim}
\usepackage{hyperref}
\usepackage{adjustbox}
\usepackage{array}

\newcommand{\abs}[1]{\left\vert #1\right\vert}

\newcommand{\ket}[1]{\left\vert{#1}\right\rangle}

\newcommand{\mean}[1]{\langle#1\rangle}

\renewcommand{\arraystretch}{1.5}

\DeclareSymbolFont{symbols}{OMS}{cmsy}{m}{n}

\begin{document}
\title{Quantum Metrology under Coarse-Grained Measurement}

\author{Byeong-Yoon Go}
\affiliation{Department of Physics, Korea Advanced Institute of Science and Technology, Daejeon, 34141, Republic of Korea}

\author{Geunhee Gwak}
\affiliation{Department of Physics, Korea Advanced Institute of Science and Technology, Daejeon, 34141, Republic of Korea}

\author{Young-Do Yoon}
\affiliation{Department of Physics, Korea Advanced Institute of Science and Technology, Daejeon, 34141, Republic of Korea}

\author{Sungho Lee}
\affiliation{Korea Astronomy and Space Science Institute, Daejeon 34055, Republic of Korea}

\author{Nicolas Treps}
\affiliation{Laboratoire Kastler Brossel, Sorbonne Universit\'{e}, ENS-Universit\'{e} PSL, CNRS, Coll\`{e}ge de France, 4 place Jussieu, F-75252 Paris, France}

\author{Jiyong Park} 
\affiliation{School of Basic Sciences, Hanbat National University, Daejeon 34158, Korea}

\author{Young-Sik Ra}
\email{youngsikra@gmail.com}
\affiliation{Department of Physics, Korea Advanced Institute of Science and Technology, Daejeon, 34141, Republic of Korea}

\date{\today}

\renewcommand{\abstractname}{Abstract}

\begin{abstract}
While quantum metrology enables measurement precision beyond classical limits, its performance is often susceptible to experimental imperfections. Most prior studies have focused on imperfections in quantum states and operations. Here, we investigate the effect of coarse graining in quantum measurement through both theoretical analysis and experimental demonstration. Using an interferometer with a squeezed vacuum and a laser input, we analyze how coarse graining in homodyne detection affects the precision of phase estimation. We evaluate the Fisher information under various coarse-graining conditions and determine, in each case, an optimal estimation strategy that saturates the Cram\'{e}r-Rao bound. Remarkably, even extremely coarse-grained measurement---with only two bins---enables phase estimation beyond the standard quantum limit and even achieves a precision that follows the Heisenberg scaling. We experimentally demonstrate quantum-enhanced phase estimation under coarse-grained homodyne detection. To determine an optimal estimation strategy, we employ the method of moments and present calibration procedures that enable its application to general experimental settings. Using only two bins, we observe a quantum enhancement of 1.2 dB compared to the classical method using the ideal measurement, improving towards 3.8 dB as the bin number increases. These results highlight a practical pathway to achieving quantum enhancement under the presence of severe experimental imperfections.
\end{abstract}

\maketitle

\section{Introduction}

Quantum metrology is a rapidly advancing field that pushes the boundaries of measurement precision by leveraging the unique properties of quantum physics~\cite{Giovannetti2004,Giovannetti2011}. Squeezed light is a key resource in this field, exhibiting nonclassical noise reduction leading to the enhancement of measurement precision~\cite{Andersen2016}. It enables phase estimation beyond the standard quantum limit (SQL)~\cite{Caves1981, Nielsen2023} and can even achieve the Heisenberg scaling~\cite{Pezze2008}. Moreover, squeezed light can be generated via deterministic optical processes~\cite{Andersen2016, Gwak2025} and is compatible with intense laser field for quantum-enhanced measurement~\cite{Caves1981}. Consequently, squeezed light has been widely employed in practical sensing applications, including gravitational wave detection~\cite{Tse2019,Acernese2019,Ganapathy2023,Acernese2023}, optomechanical sensing~\cite{Pooser2015,Xia2022}, magnetometry~\cite{Li2018}, imaging~\cite{Treps2003,Taylor2013,Casacio2021}, and spectroscopy~\cite{Herman2025,Adamou2025}.

However, such quantum enhancement is fragile under realistic experimental conditions. It is therefore crucial to systematically investigate how experimental imperfections affect the performance of quantum metrology. Along this direction, there have been extensive studies investigating optical loss~\cite{Ono2010,Oh2017,Huang2023,Gard2017}, phase diffusion~\cite{Genoni2011,Gard2017,Carrara2020}, thermal noise ~\cite{Oh2019,Oh2021}, and mode mismatch~\cite{Steinlechner2018,Roh2021}. These works essentially focus on the imperfections in quantum states and quantum operations. In contrast, imperfections in quantum measurements have remained relatively unexplored~\cite{Gessner2020,Sorelli2021,Len2022}.

Coarse graining is a common imperfection arising in realistic quantum measurement: it degrades measurement resolution by grouping nearby outcomes into the same bin~\cite{Kofler2007}. For example, in imaging, a pixelized camera sensor discretizes spatial information~\cite{Tasca2013}, and, in analog-to-digital conversion, a continuous electronic signal is digitized for readout~\cite{Gabriel:2010gb, Michel2019}.
Homodyne detection---central to quantum metrology based on squeezed light---is also subject to coarse graining due to the discretization of quadrature outcomes~\cite{Gabriel:2010gb,Park2014,Michel2019}. This effect becomes more pronounced with increasing squeezing level~\cite{Vahlbruch2016} and extending detection range~\cite{Michel2019,Ganapathy2023,Acernese2023,Herman2025}.
Coarse graining has been studied in various contexts, including the quantum-to-classical transition~\cite{Raeisi2011}, entanglement detection~\cite{Tasca2013}, and quantum steering~\cite{Schneeloch2013}. Beneficial aspects have also been reported, for example in nonlocality tests~\cite{GarciaPatron2004}, nonclassicality certification~\cite{Roh2023}, and superresolution measurements~\cite{Schaefermeier2018}. However, its effect on quantum metrology remains largely unexplored, despite its relevance to many practical sensing applications~\cite{Tse2019,Acernese2019,Ganapathy2023,Acernese2023,Pooser2015,Xia2022,Li2018,Treps2003,Taylor2013,Casacio2021,Herman2025,Adamou2025}. Note that coarse graining is a unique imperfection in measurement that cannot be addressed by existing studies on imperfections in quantum states and operations~\cite{Ono2010,Oh2017,Huang2023,Gard2017,Genoni2011,Gard2017,Carrara2020,Oh2019,Oh2021,Steinlechner2018,Roh2021}.

In this work, we present a theoretical and experimental study of quantum metrology under coarse-grained measurement. We consider a widely adopted phase-estimation scheme proposed by Caves~\cite{Caves1981}, in which a squeezed vacuum state and a coherent state are injected into a Mach-Zehnder interferometer (MZI), and homodyne detection is used to measure quadrature outcomes at an output port. The homodyne detection is subject to coarse graining, where the outcomes are discretized into finite-sized bins.

To investigate the maximum phase sensitivity achievable under coarse-grained measurement, we first evaluate the Fisher information (FI) of the measurement. Specifically, we consider two types of coarse graining, (i) binning in an equal size and (ii) binning in optimal sizes for quadrature outcomes, and vary the bin number from two to ten across the detection range. Remarkably, coarse-grained measurement retains a large fraction of FI compared to the ideal fine-grained measurement---64 \% with only two bins, which rapidly approaches the ideal value as the number of bins increases. This large fraction of FI under coarse-grained measurement not only enables quantum-enhanced phase estimation beyond the SQL, but also supports phase estimation that exhibits a Heisenberg scaling.
To determine an optimal estimation strategy that saturates the Cram\'{e}r-Rao bound (CRB), we employ the method of moments~\cite{Gessner2019,Sorelli2021b}. By formulating multiple observables corresponding to all bins in coarse-grained measurement, we derive the optimal weights for their linear combination that lead to the saturation of the CRB.

Furthermore, as a proof-of-principle experiment, we demonstrate quantum-enhanced phase estimation under coarse-grained homodyne detection. At first, with the ideal fine-grained measurement, we observe a quantum enhancement of 3.8 dB in phase estimation. Next, we investigate coarse-grained measurement with various bin numbers. To determine the optimal estimation strategy in each case, we apply the method of moments directly to the experiment. More specifically, we calibrate the actual experimental setup by measuring the covariance matrix and the mean values associated with the multiple observables of the coarse-grained measurement. Based on the calibration results, we identify the optimal combination of the observables tailored for the experimental setup and conduct the phase estimation. As a result, with just two bins, we observe a quantum enhancement of 1.2~dB over the classical method using fine-grained measurement, which rapidly improves towards the fine-grained result as the bin number increases.

\section{Theoretical framework}

\subsection{Fisher information of coarse-grained measurement}

\begin{figure}[t]
\centering
\includegraphics[width=0.48\textwidth]{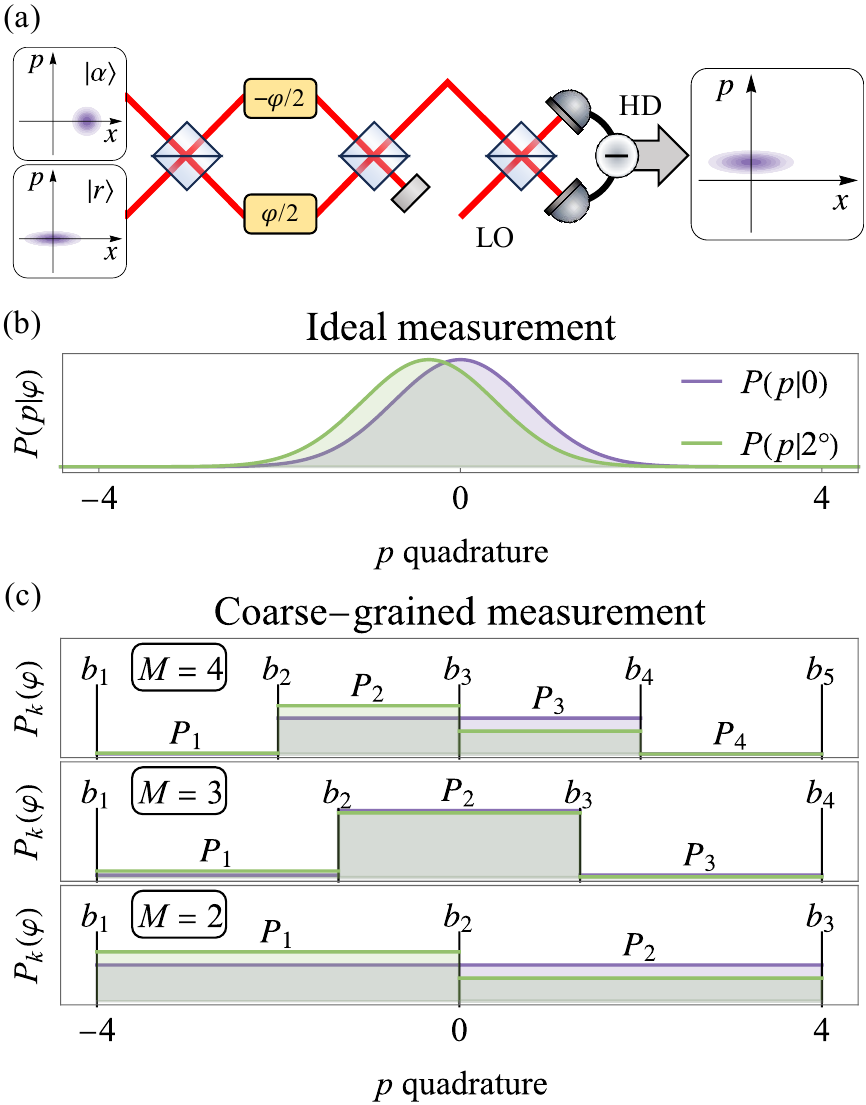}
\caption{
(a) Phase estimation with coherent state $\ket{\alpha}$ and squeezed state $\ket{r}$ input, first introduced in Ref~\cite{Caves1981}. Homodyne detection measures $\hat{p}$ quadrature at the dark port. $\alpha$: coherent-state amplitude, $r$: squeezing parameter, HD: homodyne detection, LO: local oscillator. (b) Probability density function of a quadrature outcome under ideal fine-grained measurement, given by Eq.~(\ref{eq:Gprob}). (c) Detection probability in $k$-th bin under coarse-grained measurements ($M$: bin number), given by Eq.~(\ref{eq:meano}). Both plots show an example of $\alpha=10$ and $r=0.3$ within the quadrature range of $\abs{p} \leq 4$. The green and purple curves show the distributions at $\varphi=0$ and $\varphi=2^{\circ}$, respectively. The bin boundary is denoted by $b_{k}$.
}
\label{fig:fig1}
\end{figure}

We consider the phase estimation scheme shown in Fig.~\ref{fig:fig1}(a), where a coherent state $\ket{\alpha}$ ($\alpha > 0$) and a $\hat{p}$-quadrature squeezed vacuum state $\ket{r}$ are injected into a MZI with phase difference $\varphi$ between the two arms. The quantum Fisher information (QFI) in this case is given by $\alpha^{2}e^{2r}+\sinh^{2}{r}$~\cite{Pezze2008,Ataman2019}, and in a practical case using a strong amplitude ($\alpha^{2} \gg \sinh^{2}{r}$) and a small phase shift $\varphi \ll 1$~\cite{Tse2019,Acernese2019,Ganapathy2023,Acernese2023,Pooser2015,Xia2022,Li2018,Treps2003,Taylor2013,Casacio2021,Herman2025,Adamou2025}, QFI becomes $\alpha^{2}e^{2r}$, and homodyne detection offers a phase estimation error that approaches the quantum Cram\'{e}r-Rao bound, $\Delta\varphi^{2} \propto e^{-2r}/\alpha^{2}$~\cite{Pezze2008,Ataman2019}.

At the dark output port of the MZI, the probability density function of $\hat{p}$-quadrature measurement for a given phase $\varphi$ exhibits a Gaussian distribution:
\begin{align}
    P(p|\varphi)=\frac{1}{\sqrt{2\pi}\sigma(\varphi)}\text{exp}\left[-\left( \frac{p-\bar{p}(\varphi)}{\sqrt{2}\sigma(\varphi)} \right)^{2}\right].
    \label{eq:Gprob}
\end{align}
The mean and variance of the distribution are given by $\bar{p}(\varphi)=-2\alpha\sin{(\varphi/2)}$ and $\sigma(\varphi)^{2}=\sin^{2}{(\varphi/2)}+e^{-2r}\cos^{2}{(\varphi/2)}$, respectively. Here we use the convention of $\hat{x}=\hat{a}+\hat{a}^{\dagger}$ and $\hat{p}=-i(\hat{a}-\hat{a}^{\dagger})$ ($\hat{a}$: annihilation operator, $\hat{a}^\dagger$: creation operator), resulting in the commutation relation of $[\hat{x},\hat{p}]=2i$. Fig.~\ref{fig:fig1}(b) illustrates an example of $\hat{p}$-quadrature measurement with $\alpha=10$ and $r=0.3$. A small phase shift $\varphi$ induces a change in $\bar{p}$ with a negligible effect on $\sigma^{2}$, enabling the estimation of $\varphi$ via the relation of $\varphi= - 2 \sin^{-1}(\bar{p}/2\alpha)$~\cite{Caves1981}.

To take into account realistic quantum measurement, we now investigate coarse-grained homodyne detection. It is modeled as $M$ number of discretized bins
over a finite range of $\hat{p}$-quadrature measurement $\abs{p} \leq R$. Figure~\ref{fig:fig1}(c) shows examples, where $k^{\text{th}}$ bin has boundaries of $b_{k}$ and $b_{k+1}$ for $k \in \{1,2,\ldots,M\}$. 
The detection probability in $k^{\text{th}}$ bin, $P_{k}(\varphi)$, is given by the integration of Eq.~(\ref{eq:Gprob}) over $p$ between the boundaries $b_{k}$ and $b_{k+1}$,
\begin{align}
    P_{k}(\varphi)=\frac{1}{2}\left[\text{erf}\left(\frac{b_{k+1}-\bar{p}(\varphi)}{\sqrt{2}\sigma(\varphi)}\right)-\text{erf}\left(\frac{b_{k}-\bar{p}(\varphi)}{\sqrt{2}\sigma(\varphi)}\right)\right].
\label{eq:meano}
\end{align}
Figure~\ref{fig:fig1}(c) illustrates the detection probabilities of coarse-grained measurement using equal-sized bins for various bin numbers, $M=2,3,4$. A phase shift $\varphi$ redistributes the detection probabilities among $M$ bins, inducing slight changes in $P_{k}$. However, unlike the fine-grained case in Eq. (\ref{eq:Gprob}), estimating $\varphi$ from $P_{k}(\varphi)$ requires a more advanced method.

Let us determine FI to obtain the lower bound on the phase estimation error $\Delta\varphi^{2}$ via the Cram\'{e}r-Rao bound:
\begin{align}
    \Delta\varphi^{2} \ge \frac{1}{\nu F(\varphi)},
\label{eq:CRB}
\end{align}
where $F(\varphi)$ is FI~\cite{Barbieri:2022hq,Polino2020}, $\nu$ is the number of repeated measurements, and the inequality can be saturated by employing an optimal estimation strategy~\cite{Barbieri:2022hq,Polino2020}. FI of the ideal homodyne detection involving a Gaussian state is determined by the squares of phase derivative of the mean value $(\partial_{\varphi}\bar{p})^{2}$ and standard deviation $(\partial_{\varphi}\sigma)^{2}$~\cite{Oh2019,Pinel2012}. For a phase estimation close to $\varphi=0$, the contribution of $\partial_{\varphi}\sigma$ becomes negligible compared to that of the mean value, and FI becomes~\cite{Oh2019,Pinel2012}
\begin{align}
    F_{\text{id}}(0) &= \frac{1}{\sigma^{2}}\left( \frac{\partial \bar{p}}{\partial\varphi} \right)^{2} = \alpha^{2}e^{2r}.
\label{eq:FIx}
\end{align}
As expected, $F_{\text{id}}(0)$ coincides with the QFI in the strong amplitude limit. FI for the coarse-grained measurement is given by
\begin{align}
    F_M(\varphi)=\sum^{M}_{k=1}{\frac{1}{P_{k}(\varphi)}}\left( \frac{\partial P_{k}(\varphi)}{\partial\varphi} \right)^{2},
\label{eq:FI}
\end{align}
where $P_{k}(\varphi)$ is in Eq.~(\ref{eq:meano}). To compare the FI under coarse-grained measurement (Eq.~(\ref{eq:FI})) with that of the ideal measurement (Eq.~(\ref{eq:FIx})) at $\varphi=0$, we define the Fisher information ratio $f_M$:
\begin{align}
    f_M =\frac{F_M(0)}{F_{\text{id}}(0)}=\frac{1}{\pi}\sum_{k=1}^{M}{\frac{\left( e^{-c_{k+1}^{2}}-e^{-c_{k}^{2}} \right)^{2}}{\text{erf}(c_{k+1})-\text{erf}(c_{k})}},
\label{eq:FI/FIx}
\end{align}
which depends solely on $c_{k} = e^{r}b_{k}/\sqrt{2}$. This ratio quantifies the fraction of the ideal FI that remains accessible to coarse-grained measurement. We consider two representative coarse-grained measurement configurations for a bin number $M$: equal-sized bins and optimized bins. In the former, all $M$ bins have an equal size (i.e. $b_{k+1}-b_{k}=2R/M~\forall k$). In the latter, the bin sizes are optimized to maximize the FI ratio in Eq.~(\ref{eq:FI/FIx}), giving the tight upper bound on the achievable FI ratio. Throughout our analysis, we consider the four-standard-deviation range $R=4\sigma(0)=4 e^{-r}$, which captures nearly all events except about 63 ppm. Note that $b_k$ scales linearly with $R=4 e^{-r}$, leading to the cancellation of $e^r$ in $c_k$. As a result, the FI ratio becomes independent of the squeezing parameter $r$.

\begin{figure}[t]
\centering
\includegraphics[width=0.48\textwidth]{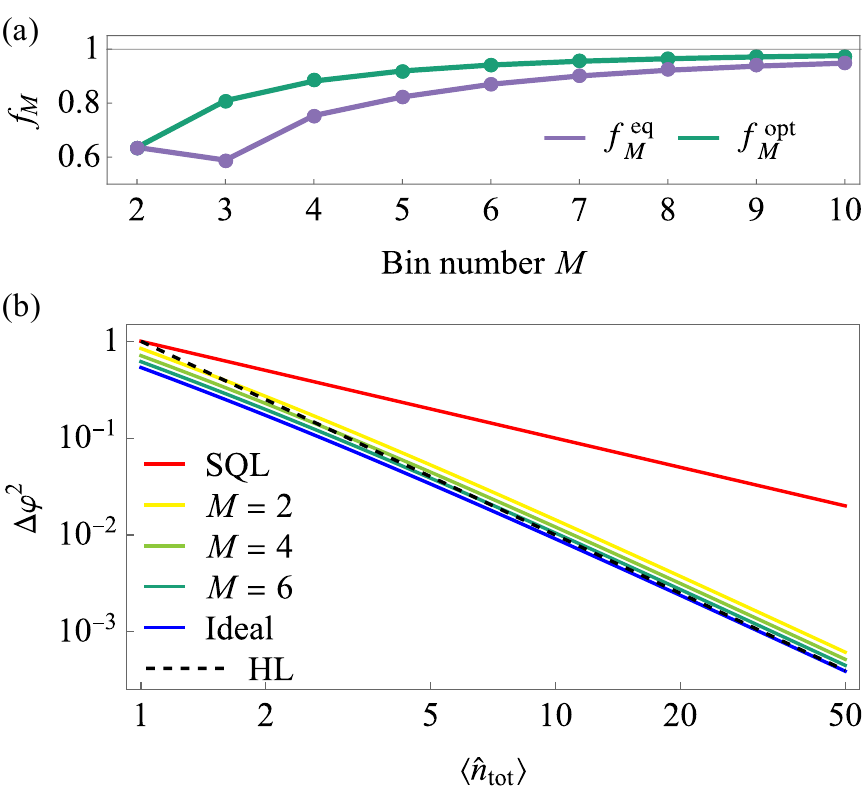}
\caption{
(a) Fisher information ratio $f_M=F_M(0)/F_{\text{id}}(0)$ in Eq.~(\ref{eq:FI/FIx}) as a function of the bin number $M$. In the analysis, we consider the total range of the four-standard-deviation of a quadrature distribution, as explained in the main text. 
The purple and green dots represent the Fisher information ratio under the equal-sized bins ($f^{\text{eq}}_M$) and optimal bins ($f^{\text{opt}}_M$), respectively.
(b) Phase estimation error as increasing the total average photon number $\mean{\hat{n}_{\text{tot}}}$. We consider the equal distribution of the total photons between the squeezed vacuum and the coherent state, which gives a Heisenberg scaling in the case of ideal homodyne detection~\cite{Pezze2008}. HL and SQL represent the Heisenberg limit ($\Delta\varphi = \mean{\hat{n}_{\text{tot}}}^{-1}$, the black dashed line) and the standard quantum limit ($\Delta\varphi = \mean{\hat{n}_{\text{tot}}}^{-1/2}$, the red solid line), respectively. Coarse-grained measurement for a bin number $M$ exhibits only a vertical shift from the ideal homodyne detection, thus exhibiting the Heisenberg scaling as well.
}
\label{fig:fig2}
\end{figure}

Figure~\ref{fig:fig2}(a) shows the FI ratio $f_M$ in Eq.~(\ref{eq:FI/FIx}) for the two configurations.
Under the extremely coarse-grained measurement with $M=2$, where the bin boundaries are given by $-b_{1}=b_{3}=R$ and $b_{2}=0$, the FI ratio already amounts to about \( 64\% \).
The FI ratio under optimal binning (\( f^{\text{opt}}_M \)) increases monotonically with \( M \), as each additional bin boundary introduces an extra parameter to optimize the FI ratio. On the other hand, the FI ratio with equal bins (\( f^{\text{eq}}_M \)) does not necessarily increase with \( M \), because the bin boundaries, constrained by the total range \( R \), can be substantially deviated from the optimal condition (see $M=3$ in Fig.~\ref{fig:fig2}(a)).
Nevertheless, the FI ratio of $M$ bins is always greater than that of any divisor of \( M \), since the former includes all boundaries of the latter. In both coarse-grained configurations, we observe that the FI ratio \( f_M \) rapidly approaches unity with increasing $M$, attaining \( f^{\text{eq}}_{10} \simeq 95\% \) and \( f^{\text{opt}}_{10} \simeq 98\% \).

These remarkably high FI ratios enable quantum-enhanced phase estimation under coarse-grained homodyne detection. For the case of two bins, the reduction in FI (to $64 \%$ of the ideal measurement) caused by coarse-graining can be fully compensated by introducing more than 2 dB of squeezing. At an even higher squeezing level, quantum-enhanced phase estimation can be achieved in the presence of the coarse-graining effect, surpassing the standard quantum limit (SQL) by the ideal fine-grained measurement. Furthermore, coarse-grained measurement can even achieve phase estimation exhibiting a Heisenberg scaling. In the ideal homodyne detection, Heisenberg scaling is achieved by equally distributing the total mean photon number $\mean{\hat{n}_{\text{tot}}}$ between the squeezed vacuum and the coherent state~\cite{Pezze2008}, i.e., \( \sinh^{2}{r} = \alpha^{2} = \mean{\hat{n}_{\text{tot}}}/2 \), yielding FI of $\alpha^{2}e^{2r}$ at $\varphi=0$. The coarse graining merely introduces a constant multiplication factor, $f_M$, to this FI, resulting in $f_M \alpha^{2}e^{2r}$, and thus, the Heisenberg scaling is still attained. Figure~\ref{fig:fig2}(b) plots the phase estimation error for coarse-grained measurement, in comparison with the Heisenberg limit and the SQL. As expected, the phase estimation error under ideal measurement (blue solid line) follows the Heisenberg scaling. 
Notably, coarse-grained measurements also exhibit Heisenberg scaling with just a vertical shift with respect to the ideal case, reflecting the prefactor $f_M$ due to the coarse graining.

\subsection{Optimal phase estimation strategy}

\begin{figure}[t]
\centering
\includegraphics[width=0.48\textwidth]{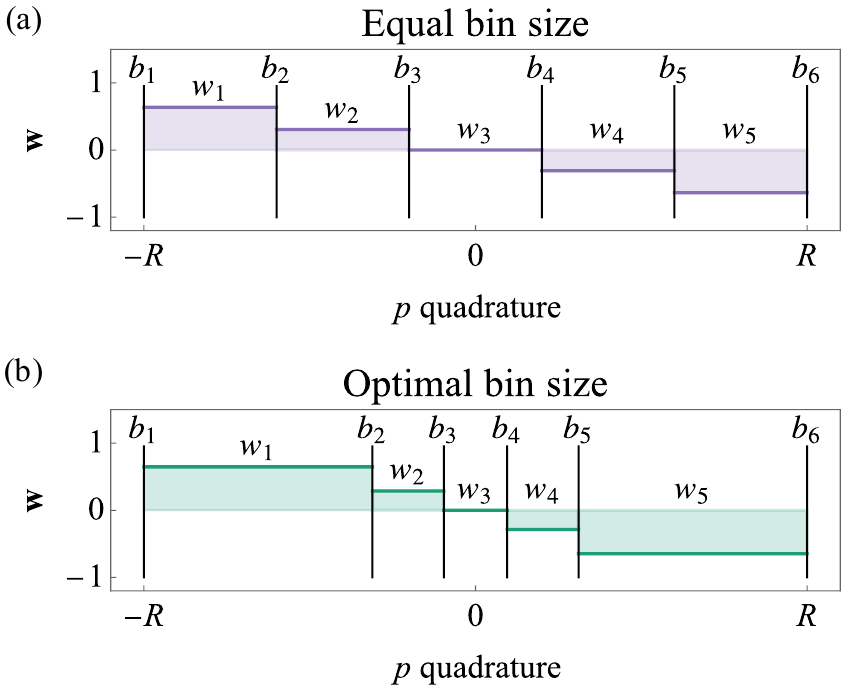}
\caption{
Examples of optimal weights for $M=5$ and $\varphi = 0$ under coarse-grained measurement with (a) equal-sized bins and (b) optimal bins, obtained from Eq.~(\ref{eq:optw}). The weight for $k^{\text{th}}$ bin (between $b_{k}$ and $b_{k+1}$) is denoted by $w_{k}$.
}
\label{fig:fig3}
\end{figure}

Let us now determine the optimal strategy for phase estimation under coarse-grained measurement by saturating the CRB using the FI in Eq.~(\ref{eq:FI}). First, for the ideal homodyne detection, it is well known that the mean quadrature value, $\bar{p}$, serves as an optimal estimator, yielding a phase estimation error $\Delta\varphi_{\text{id}}$ that saturates the CRB using Eq. (\ref{eq:FIx})~\cite{Pinel2012,Ataman2019},
\begin{align}
    \Delta\varphi_{\text{id}}^{2}=\frac{\sigma^{2}}{\nu \abs{\frac{\partial\bar{p}}{\partial\varphi}}^{2}}=\frac{1}{\nu F_{\text{id}}(0)}.
\label{eq:phi0}
\end{align}
In contrast, coarse-grained measurement requires a more general approach which can account for discretized observables. Here we use the method of moments, which employs the mean values of multiple observables, to determine the optimal linear combination of the observables associated with coarse-grained measurement~\cite{Sorelli2021,Gessner2019,Sorelli2021b}.

We start by formulating an observable \( \hat{o}_{k} \) associated with the discretized measurement in \( k^{\text{th}} \) bin:
\begin{align}
    \hat{o}_{k} = \int_{b_{k}}^{b_{k+1}}dp\vert p\rangle\langle p\vert,
\end{align}
where $\hat{p} \vert p\rangle = p \vert p\rangle$. Note that \( \hat{o}_{k} \), corresponding to the projection operator onto the subspace of \( k^{\text{th}} \) bin, assigns a value of one when the quadrature outcome lies within the bin range \( [b_{k}, b_{k+1}] \) and zero otherwise. Accordingly, the expectation value \( \mean{\hat{o}_{k}} \) gives the probability that a quadrature outcome is registered in the \( k^{\text{th}} \) bin, i.e., \( \mean{\hat{o}_{k}} = P_{k} \). Note that \( \mean{\hat{o}_{k} \hat{o}_{l}} = 0 \) for \( k \neq l \), because a quadrature outcome cannot be registered in two different bins together. The covariance matrix, \( \Gamma_{kl} = \mean{\hat{o}_{k} \hat{o}_{l}} - \mean{\hat{o}_{k}} \mean{\hat{o}_{l}} \), thus evaluates to \( \Gamma_{kl} = -P_{k} P_{l} \) for \( k \neq l \), and \( \Gamma_{kl} = P_{k} (1 - P_{k} ) \) for \( k = l \). This covariance matrix reflects the expected anti-correlation $\Gamma_{kl}<0$, arising from the impossibility of simultaneous registration of a quadrature outcome in two distinct bins.

Next, we consider a linear combination of all $M$ observables $\hat{\mathbf{o}}=(\hat{o}_{1}, \hat{o}_{2}, \ldots, \hat{o}_{M})^T$ by a weight $\textbf{w} = (w_{1}, w_{2}, \ldots, w_{M})$: $\hat{O} = \textbf{w}^{T} \hat{\textbf{o}}$. The expectation value of the combined observable $\hat{O}$ is
\begin{align}
\mean{\hat{O}}_\varphi=\textbf{w}^{T}\mean{\hat{\textbf{o}}}_\varphi \equiv g(\varphi),
\label{eq:est}
\end{align}
where the dependence on $\varphi$ is explicitly shown as a subscript. $g(\varphi)$ is used as the calibration function to estimate $\varphi$ from the measurement of $\hat{O}$. More specifically, the estimator of $\varphi$, denoted by $\tilde{\varphi}$, is given by $\tilde{\varphi}=g^{-1} (\bar{O})$, with $\bar{O}$ the sample mean of $\hat{O}$ over $\nu$ repeated measurements. The error of this estimator is given by
\begin{align}
    \Delta\varphi^{2}=\frac{\mean{\Delta \hat{O}}_\varphi^{2}}{\nu\abs{\frac{\partial\mean{\hat{O}}_\varphi}{\partial\varphi}}^{2}}=\frac{\textbf{w}^{T}\mathbf{\Gamma}_{\varphi}\textbf{w}}{\nu\abs{\textbf{w}^{T}\frac{\partial\mean{\hat{\textbf{o}}}_\varphi}{\partial\varphi}}^{2}},
\label{eq:phi}
\end{align}
where $\mathbf{\Gamma}_\varphi$ is the covariance matrix defined above. The estimation error depends on the weight vector $\textbf{w}$, and the optimal weight that achieves this minimum error 
is given by
\begin{align}
    \textbf{w}_{\varphi}=
\mathbf{\Gamma}^{+}_{\varphi}\frac{\partial\mean{\hat{\textbf{o}}}_\varphi}{\partial\varphi} 
.
\label{eq:optw}
\end{align}
Here, $\mathbf{\Gamma}^{+}_\varphi$ is the pseudoinverse matrix of $\mathbf{\Gamma}_\varphi$, extending earlier results that relied on $\mathbf{\Gamma}^{-1}_\varphi$, valid only for invertible covariance matrices~\cite{Gessner2019,Sorelli2021,Sorelli2021b}. See the Appendix~\ref{Appendix:A} for the proof. With this optimal weight, the phase estimation error 
is given by
\begin{align}
    \Delta\varphi^{2}=\frac{1}{\nu\frac{\partial\mean{\hat{\textbf{o}}}_\varphi^{T}}{\partial\varphi}\mathbf{\Gamma}^{+}_{\varphi}\frac{\partial\mean{\hat{\textbf{o}}}_\varphi}{\partial\varphi}}=\frac{1}{\nu F_M(\varphi)},
\end{align}
saturating the CRB based on Eq.~(\ref{eq:FI}). Figure~\ref{fig:fig3} shows examples of the optimal weights for coarse-grained measurements: equal-sized bins in Fig.~\ref{fig:fig3}(a) and optimal bins in Fig.~\ref{fig:fig3}(b). Note that these optimal weights deviate from a simple linear trend, where the actual values are presented in the Appendix~\ref{Appendix:B} up to $M=10$. Moreover, the method used to determine the optimal weight in Eq. (\ref{eq:optw}) can be directly applied to actual experimental setups by measuring $\mathbf{\Gamma}_\varphi$ and $\mean{\hat{\textbf{o}}}_\varphi$, offering a practical means to identify optimal strategies in general experimental conditions. 

\section{Experiment}

\subsection{Experimental setup}

\begin{figure}[t]
\centering
\includegraphics[width=0.48\textwidth]{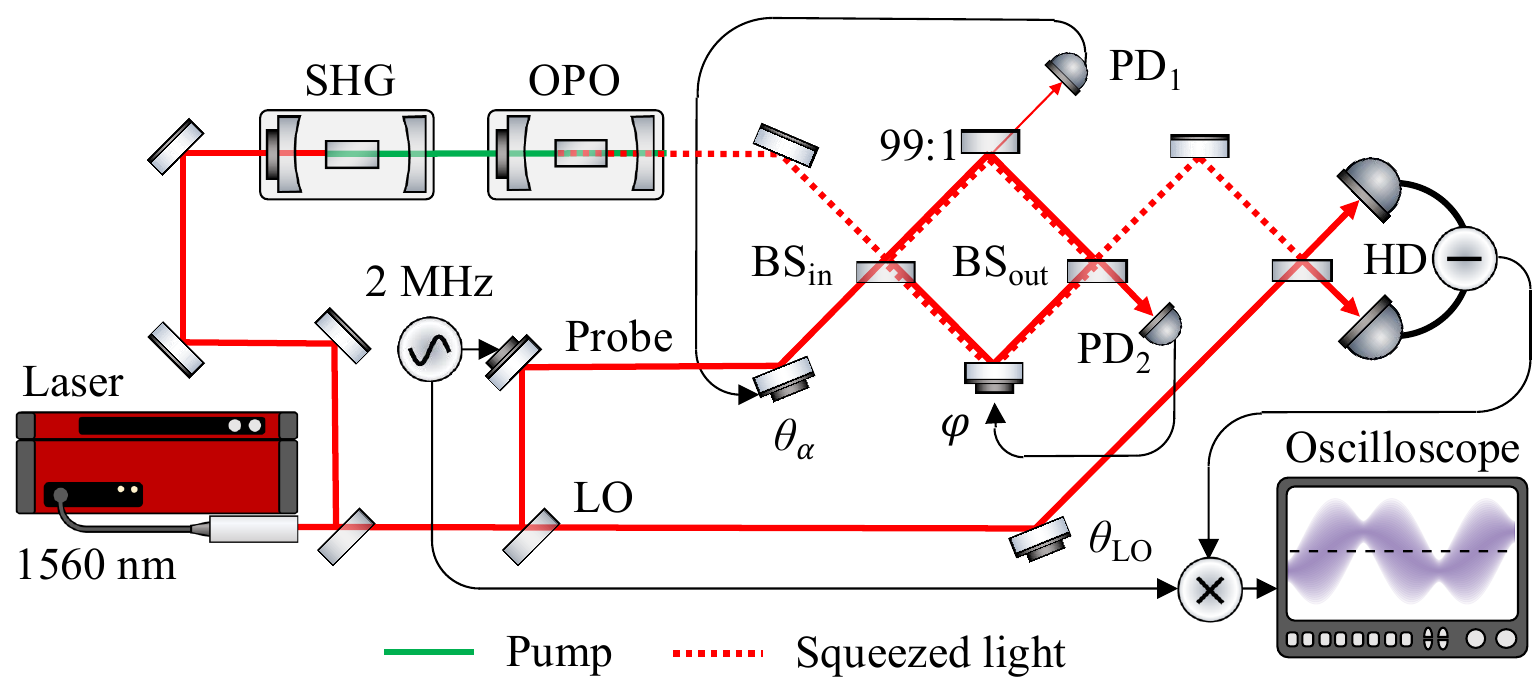}
\caption{
Experimental setup for measuring the interferometric phase $\varphi$. A squeezed vacuum state (from OPO) and a coherent state (in probe) are used as the input of the interferometer. The output from the dark port of the interferometer is detected by homodyne detector (HD). Quadrature outcomes are obtained with an oscilloscope after demodulating the HD signals at the frequency of 2~MHz. Coarse-graining effects are investigated by binning the quadrature outcomes in various conditions. SHG: second-harmonic generation, OPO: optical parametric oscillator, $\text{BS}_{\text{in(out)}}$: input(output) beam splitter, $\text{PD}_{1(2)}$: photodiode 1(2), LO: local oscillator.
}
\label{fig:fig4}
\end{figure}

We experimentally investigate quantum-enhanced phase estimation under coarse-grained homodyne detection. Figure~\ref{fig:fig4} describes the experimental setup. A 1560 nm wavelength laser is first split into two beams. One beam is frequency-doubled via cavity-enhanced second-harmonic generation (SHG) with a 10-mm periodically poled lithium niobate (PPLN) crystal. The generated 780 nm pump laser (green line) is then coupled to an optical parametric oscillator (OPO), pumping a nonlinear crystal to generate squeezed light (red dashed line). The OPO is a linear cavity with a bandwidth of 43 MHz, containing a 10 mm periodically poled potassium titanyl phosphate (PPKTP) crystal for a degenerate type-0 parametric down-conversion process. We inject a locking beam to the OPO for locking both the OPO length to maintain the resonant condition (via the Pound–Drever–Hall technique) and the pump phase to maintain the deamplification condition. Note that we omit the locking implementation in Fig.~\ref{fig:fig4} for clarity. The locking beam is phase-modulated at 30 MHz, and the photocurrent signal by measuring the reflected locking beam from the OPO is demodulated at the same frequency. Demodulation with $\pi/2$ and 0 phase delays provides the error signals for the OPO and the pump phase lockings, respectively~\cite{Mehmet2009}. We observe 5.2 dB of squeezing at the sidebands of 2 MHz at the OPO output with the pump power of 250 mW.


The other beam from the laser is further split into a probe for phase estimation and a LO for homodyne detection. The probe is phase-modulated at 2 MHz to generate a coherent state in the antisymmetric sidebands. The probe beam is then coupled with the squeezed light beam at the input beam splitter ($\text{BS}_{\text{in}}$). To lock the phase between the two beams, a small fraction of a beam after $\text{BS}_{\text{in}}$ is tapped off using a 99:1 BS, where the photodiode ($\text{PD}_{1}$) detects the interference between the DC mean fields contained in the two input beams. To obtain the error signal, we employ a dithering locking at 1.5 kHz frequency, maintaining the constructive interference in DC. As a result, we have the input coherent state and the squeezed vacuum as described in Fig.~\ref{fig:fig1}(a), in the antisymmetric sidebands of 2 MHz. To stabilize the interferometric phase $\varphi$, we use the interference signal from the second photodiode ($\text{PD}_{2}$) through a dithering locking at 2 kHz.



The beam from the detection port of the MZI is measured by homodyne detection (HD), with a high mode-matching contrast of $98\%$ with the LO. The phase of the LO, $\theta_{\text{LO}}$, is adjusted by scanning a mirror that reflects the LO beam. The HD signal is demodulated at 2 MHz and recorded on an oscilloscope, yielding the full phase-dependent quadrature data. For each data acquisition, 100,000 data points are sampled while scanning the LO phase over approximately $2\pi$. These data points are then divided sequentially into groups containing 2,000 samples. In each group, the first and last 500 samples are used to estimate the LO phase $\theta_{\text{LO}}$, and the remaining 1,000 samples are taken as the quadrature outcomes at that phase. For the MZI phase of $\varphi = 0$, where the squeezed light is completely transferred to the HD, we observe 3.8 dB of squeezing and 15 dB of anti-squeezing in the 2 MHz sidebands. We further calibrate the coherent-state amplitude $\alpha$ in the 2 MHz sidebands of the probe by locking the MZI phase at $\varphi = \pi$ which sends the input coherent state to the HD. By measuring the mean value of the HD signal, $\bar{q}$, as a function of $\theta_{\text{LO}}$, we determine $\alpha = 5.7$ using the calibration curve $\bar{q}=2\alpha\cos \theta_{\text{LO}}$.

\subsection{Experimental results}

\begin{figure*}[htbp]
    \centering
    \makebox[\textwidth][c]{
        \includegraphics[width=\textwidth]{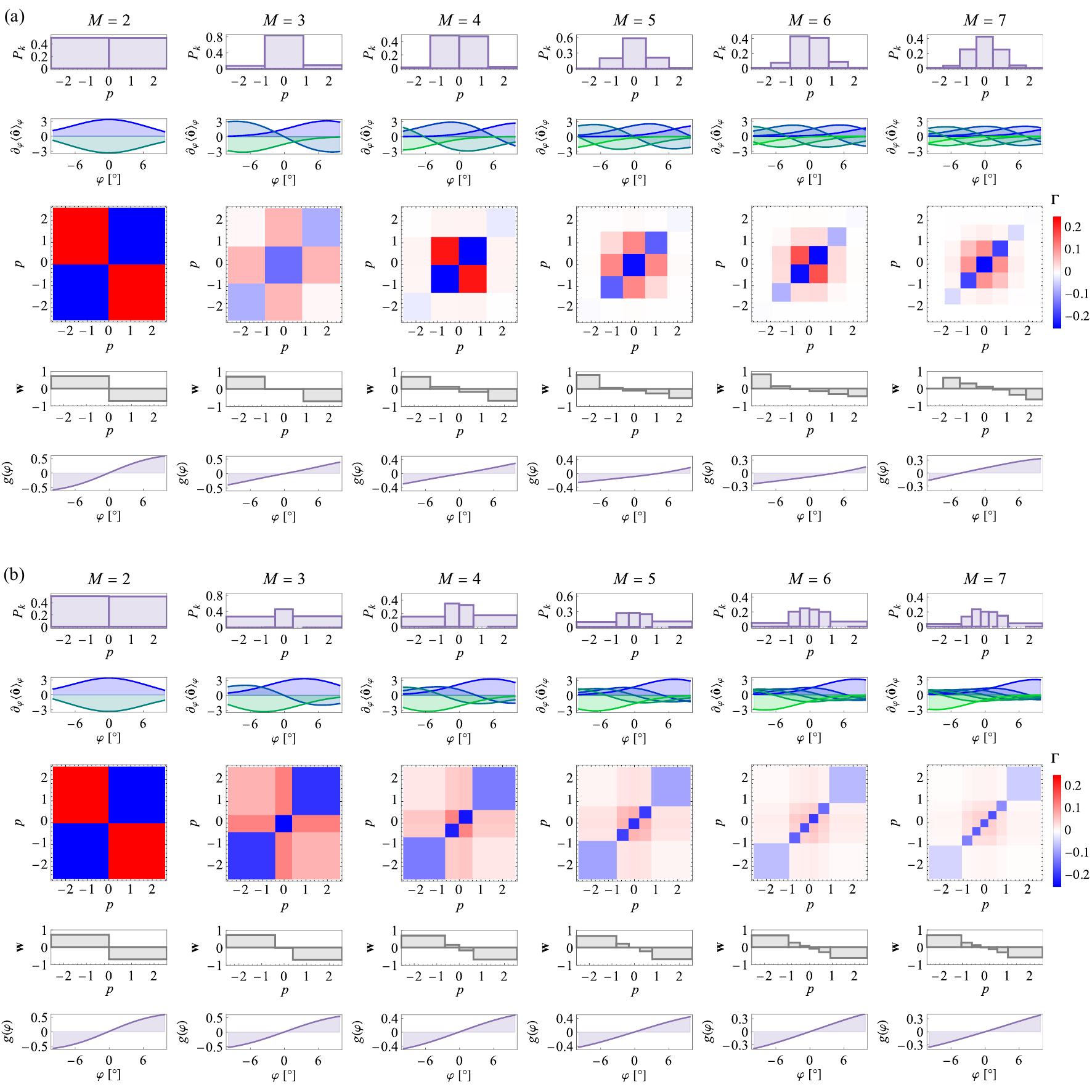}
    }
\caption{
Experimental determination of the optimal weights and the associated calibration functions. (a) Equal-sized binning and (b) optimal binning, where the panels from left to right correspond to bin numbers \( M = 2 \) to \( M = 7 \). For both (a) and (b), the first row shows the detection probabilities $P_k$ ($k=1,\dots,M$, with bins indexed from left to right bars) at the phase $\varphi_{0}=-0.02^{\circ}$; the second shows the partial derivatives of expectation values of multiple observables $\partial_{\varphi} \mean{\hat{\mathbf{o}}}_\varphi$, with the increasing bin index illustrated by a gradual color change from blue to green; the third shows the covariance matrices \( \boldsymbol{\Gamma}_{\varphi_{0}} \); the fourth shows the optimal weight vectors $\textbf{w}_{\varphi_{0}}$; and the last row shows the calibration functions $g(\varphi)$.
}
\label{fig:fig5}
\end{figure*}

Figure~\ref{fig:fig5} shows the experimental results used for determining the optimal weight $\mathbf{w}_{\varphi}$ and the associated calibration function $g(\varphi)$. Figure~\ref{fig:fig5}(a) and (b) correspond to equal-sized binning and optimal binning, respectively. We first measure the detection probabilities $P_k(\varphi)$ at various phases $\varphi$. Specifically, we record $P_k(\varphi)$ at 150 different phases in $[-20^{\circ},20^{\circ}]$; the first rows of Fig.~\ref{fig:fig5}(a) and (b) show examples of $P_k(\varphi)$ taken at \( \varphi = -0.02^{\circ} \). We then determine $\mean{\hat{\mathbf{o}}}_\varphi$ for arbitrary $\varphi$ by fitting the measured probabilities to a theoretical model in Eq. (\ref{eq:meano}). Next, a partial derivative of $\mean{\hat{\mathbf{o}}}_\varphi$ is calculated, as shown in the second rows. We further obtain a covariance matrix \( \boldsymbol{\Gamma}_{\varphi_0} \), where $\varphi_0$ is the phase that the optimal observable is constructed; in our case, $\varphi_0 = -0.02^{\circ}$, and the corresponding \( \boldsymbol{\Gamma}_{\varphi_0} \) is shown in the third row. Using the results in the second and the third rows, the optimal weight vector $\textbf{w}_{\varphi_0}$ is determined by $\mathbf{\Gamma}^{+}_{\varphi_0}~ \partial_{\varphi}\mean{\hat{\mathbf{o}}}_\varphi |_{\varphi=\varphi_0}$ (from Eq. (\ref{eq:optw})), which is shown in the fourth row. Finally, the calibration function $g(\varphi)$ is evaluated as $\textbf{w}_{\varphi_0}^{T}\mean{\hat{\textbf{o}}}_\varphi$ (from Eq. (\ref{eq:est})), enabling the phase estimation by $\tilde{\varphi}=g^{-1} (\bar{O})$, with $\bar{O}$ the sample mean of $\hat{O}=\textbf{w}^{T}_{\varphi_0} \hat{\textbf{o}}$ over $\nu$ trials.

\begin{figure}
\centering
\includegraphics[width=0.48\textwidth]{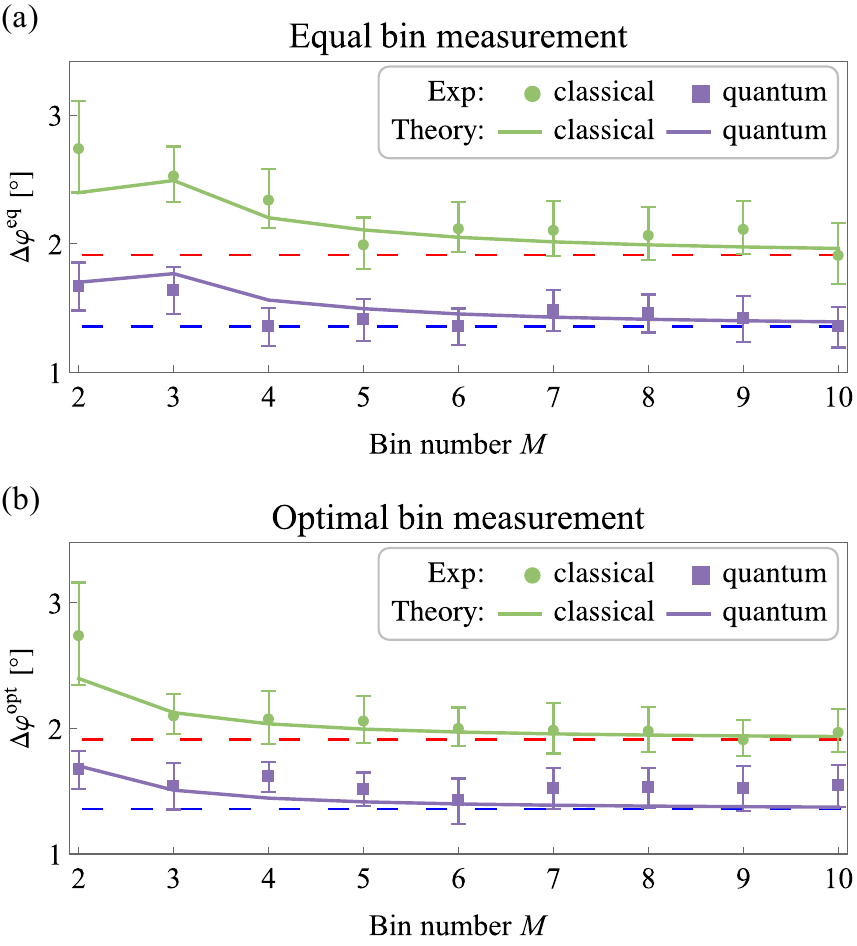}
\caption{
Phase estimation error $\Delta\varphi$ as a function of bin number \( M \). (a) Equal-sized binning and (b) optimal binning. Phase estimation is investigated at $\varphi_{0}=-0.02^{\circ}$, where the sample mean over $\nu=25$ repetitions is used, and the estimation error is evaluated as the standard deviation of 40 repeated experiments. The purple and light-green dots show experimental data obtained with squeezing (quantum) and without squeezing (classical), respectively, while the solid lines indicate the corresponding theoretical predictions. The blue and red dashed lines indicate the ideal homodyne detection with and without squeezed light, respectively. Error bars of $\Delta\varphi$ are estimated using a bootstrap method by generating 40 resampled datasets.
}
\label{fig:fig6}
\end{figure}

\begin{figure*}
\centering
\includegraphics[width=\textwidth]{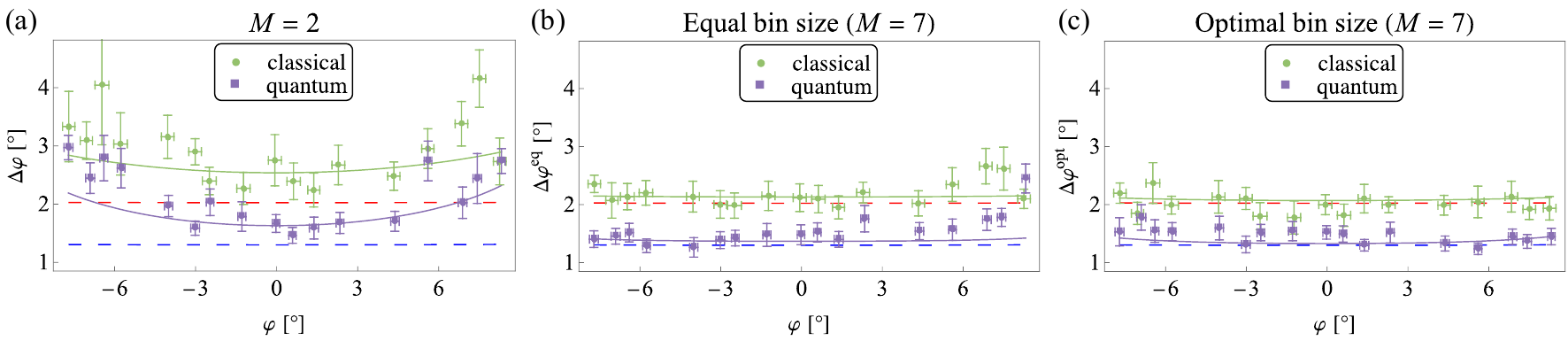}
\caption{
Phase estimation error $\Delta\varphi$ as a function of phase $\varphi$ under coarse-grained homodyne detection. For phase estimation, we use the observable optimized at $\varphi_{0}=-0.02^{\circ}$, and used the sample mean over $\nu=25$ repetitions. The estimation error is evaluated as the standard deviation of 40 repeated experiments. The purple and light green dots represent the experimentally obtained phase errors with squeezing (quantum) and without squeezing (classical), respectively, while the solid lines indicate the corresponding theoretical predictions. The blue and red dashed lines indicate the ideal homodyne measurement with and without squeezed light, respectively.
The estimation errors are shown for (a) $M=2$, (b) $M=7$ with equal-sized binning, and (c) $M=7$ with optimal binning. Error bars of $\Delta\varphi$ are estimated using a bootstrap method by generating 40 resampled datasets.
}
\label{fig:fig7}
\end{figure*}

After the calibration, we investigate the phase error $\Delta\varphi$ of the estimator $\tilde{\varphi}=g^{-1} (\bar{O})$ for $\nu=25$. Figure~\ref{fig:fig6} presents phase estimation errors at \( \varphi_{0} = -0.02^{\circ} \) for various bin numbers \( M \): (a) equal-sized binning and (b) optimal binning. As $M$ increases, the phase error under coarse-grained measurement rapidly approaches the phase error of the ideal measurement. For all bin numbers, coarse-grained measurement with squeezed light exhibits lower estimation error than the ideal measurement without squeezing, demonstrating quantum-enhanced phase estimation under coarse-grained measurement. The observed enhancement in terms of error variance $\Delta\varphi^2$ is 1.2 dB at $M=2$, and 1.4 dB at $M=3$ with equal-sized binning and 1.9 dB at $M=3$ with optimal binning. The theoretical graphs assume $\alpha=5.7$ with 3.8 dB squeezing, agreeing well with the experimental data.

Furthermore, we investigate the phase range over which an estimator optimized at a specific phase remains applicable. Figure~\ref{fig:fig7}(a) presents the results obtained by the estimator optimized at $\varphi_{0} = -0.02^{\circ}$ for $M=2$. Notably, quantum enhancement is still observed within the range \( |\varphi| \leq 6.7^{\circ} \). As the phase increases, however, the estimation error exceeds that of ideal classical homodyne detection because the Gaussian distribution in Eq. (\ref{eq:Gprob}) increasingly falls within a single bin, leading to saturation of the detection probabilities in the two bins. This effect is more detrimental in the quantum case than in the classical one, as seen by the faster increase of the estimation error in Fig.~\ref{fig:fig7}(a). This is because the quantum distribution is narrower and therefore more susceptible to such saturation.
Increasing the number of bins \( M \) further reduces the phase estimation error and also broadens the phase range in which quantum enhancement can be observed. Figure~\ref{fig:fig7}(b) shows the result for \( M = 7 \) under equal-sized configuration. The phase error at large $\varphi$ is greatly reduced compared to \( M = 2 \), extending the phase range of quantum-enhanced phase estimation. In Fig.~\ref{fig:fig7}(c), optimizing the bin sizes reduces the phase error compared to the equal-sized configuration, enhancing the phase sensitivity near $\varphi=0$. 

\section{Conclusion}
Coarse graining is a common imperfection in realistic quantum measurements that substantially limits measurement resolution. By overcoming this limitation, we introduce an optimal strategy for phase estimation under coarse-grained measurements. 
Using the Fisher information, we analyze the fundamental capability of coarse-grained measurement for phase estimation and find that it can retain a remarkably large fraction of the Fisher information of the ideal measurement. Consequently, coarse-grained measurement can still achieve quantum-enhanced and even Heisenberg-limited phase estimation. To fully leverage this potential, we provide a method to find an optimal observable for phase estimation that saturates the Cram\'{e}r-Rao bound. 
In a proof-of-principle experiment using 3.8 dB of squeezed vacuum, we observe a quantum enhancement of 1.2 dB even with a two-bin measurement. We further find that increasing the number of bins rapidly improves the enhancement toward the ideal limit and extends the phase range over which quantum enhancement can be attained. This technique will become increasingly critical as squeezing levels rise~\cite{Vahlbruch2016} and homodyne detection ranges expand~\cite{Michel2019,Ganapathy2023,Acernese2023,Herman2025}. Moreover, our approach can be further applied to a wide range of sensing applications based on continuous-variable quantum measurements, such as quantum imaging~\cite{Treps2003,Taylor2013,Casacio2021} and quantum spectroscopy~\cite{Herman2025,Adamou2025}. This work provides a practical method for realizing quantum metrology in the presence of strong experimental imperfections.

\section*{Acknowledgments}

This work was supported by the Korea Astronomy and Space Science Institute under the R\&D program (Project No. 2025-1-810-00) and the Ministry of Science and ICT (MSIT) of Korea (RS-2025-00562372, RS-2024-00408271, RS-2024-00442762, RS-2023-NR119925) under the Information Technology Research Center (ITRC) support program (IITP-2026-RS-2020-II201606) and Institute of Information \& Communications Technology Planning \& Evaluation (IITP) grant (RS-2025-25464959, RS-2022-II221029).


\appendix
\section{Determining the optimal weight using the method of moments}
\label{Appendix:A}
In this section, we show the detailed derivation of the optimal weight vector given by Eq.~(\ref{eq:optw}), using the method of moments. The goal of this optimization is to find the optimal weight vector $\textbf{w}$ that minimizes the phase variance
\begin{equation}
    \Delta\varphi^{2}=\frac{\textbf{w}^{T}\mathbf{\Gamma}\textbf{w}}{\nu \textbf{w}^{T}\mathbf{\Lambda}\textbf{w}},
    \label{eq:pvar}
\end{equation}
with $\mathbf{\Lambda}=\partial_{\varphi}\mean{\hat{\textbf{o}}}_{\varphi}\partial_{\varphi}\mean{\hat{\textbf{o}}}_{\varphi}^{T}$, over the domain $\textbf{w}^{T}\mathbf{\Lambda}\textbf{w}>0$. This problem corresponds to the generalized Rayleigh quotient problem, which can be recast as the generalized eigenvalue problem,
\begin{equation}
    \mathbf{\Gamma w}=\lambda\mathbf{\Lambda w},
    \label{eq:gevp}
\end{equation}
where $\lambda$ and $\mathbf{w}$ are the associated eigenvalue and eigenvector, respectively. Substituting this relation into Eq.~(\ref{eq:pvar}) yields $\Delta\varphi^{2}=\lambda/\nu$, indicating that the smallest eigenvalue leads to the minimal phase variance.

For an invertible covariance matrix $\mathbf{\Gamma}$, Eq. (\ref{eq:gevp}) can be converted into the conventional eigenvalue problem:
\begin{equation}
    \mathbf{\Gamma}^{-1} \mathbf{\Lambda w}=\lambda^{-1} \mathbf{w}.
    \label{eq:cevp}
\end{equation}
By solving this, one obtains the eigenvalue of $\lambda^{-1}=\partial_{\varphi}\mean{\hat{\textbf{o}}}_{\varphi}^{T}\mathbf{\Gamma}^{-1}\partial_{\varphi}\mean{\hat{\textbf{o}}}_{\varphi}$ and the eigenvector of
\begin{equation}
    \textbf{w}=\mathbf{\Gamma}^{-1}\frac{\partial\mean{\hat{\textbf{o}}}_{\varphi}}{\partial\varphi},
    \label{eq:invweigth}
\end{equation}
agreeing with the optimal weight derived for an invertible matrix. Note that for a finite quadrature range of $R$, the covariance matrix $\mathbf{\Gamma}$ is invertible, and Eq. (\ref{eq:invweigth}) becomes the optimal weight.


However, $\mathbf{\Gamma}$ is non-invertible if (1) $R=\infty$, (2) $R$ is finite but sufficiently large that, when measuring the covariance matrix, no quadrature outcome $p$ that satisfies $|p| > R$ is observed. For a non-invertible $\mathbf{\Gamma}$, Eq. (\ref{eq:invweigth}) is no longer valid, and we directly address the generalized eigenvalue problem in Eq. (\ref{eq:gevp}). First, note that Eq. (\ref{eq:gevp}) leads to $\mathbf{\Gamma}\textbf{w} \propto \partial_{\varphi}\mean{\hat{\textbf{o}}}_{\varphi}$. It means that $\partial_{\varphi}\mean{\hat{\textbf{o}}}_{\varphi}$ lies within the column space of $\mathbf{\Gamma}$, which is equivalent to the following expression:
\begin{equation}
   \mathbf{\Gamma}\mathbf{\Gamma}^{+}\frac{\partial\mean{\hat{\textbf{o}}}_{\varphi}}{\partial\varphi}= \frac{\partial\mean{\hat{\textbf{o}}}_{\varphi}}{\partial\varphi},
    \label{eq:picon}
\end{equation}
where $\mathbf{\Gamma}^{+}$ is the pseudoinverse matrix of $\mathbf{\Gamma}$. This is because $ \mathbf{\Gamma}\mathbf{\Gamma}^{+}$ is a projector onto the column space of $\mathbf{\Gamma}$. In this case, $\textbf{w}$ can be written as,
\begin{equation}
  \textbf{w}=\mathbf{\Gamma}^{+}\frac{\partial\mean{\hat{\textbf{o}}}_{\varphi}}{\partial\varphi}+\textbf{v},
\end{equation}
where $\textbf{v}$ is an arbitrary vector in the kernel of $\mathbf{\Gamma}$, satisfying $\mathbf{\Gamma v}=\mathbf{0}$. 
The corresponding generalized eigenvalue in Eq.~(\ref{eq:gevp}) is then given by $\lambda^{-1}=\partial_{\varphi}\mean{\hat{\textbf{o}}}_{\varphi}^{T}\mathbf{\Gamma}^{+}\partial_{\varphi}\mean{\hat{\textbf{o}}}_{\varphi}$. 
By choosing $\mathbf{v}=\mathbf{0}$, the expression for the optimal weight vector simplifies to the form given in Eq.~(\ref{eq:optw}).

In the coarse-grained homodyne detection, the pseudoinverse of the covariance matrix is given by
\begin{equation}
    (\mathbf{\Gamma}^{+})_{kl}=\frac{\delta_{kl}}{P_{k}}-\frac{1}{M}\left(\frac{1}{P_{k}}+\frac{1}{P_{l}} \right)+\frac{1}{M^{2}}\sum^{M}_{m=1}\frac{1}{P_{m}}.
\end{equation}
It satisfies the following relations:
\begin{equation}
    \mathbf{\Gamma}\mathbf{\Gamma}^{+}=\mathbf{\Gamma}^{+}\mathbf{\Gamma}=\mathbf{I}_{M}-\frac{1}{M}\mathbf{J}_{M},
    \label{eq:piid}
\end{equation}
where $\mathbf{I}_{M}$ is the $M\times M$ identity matrix and $\mathbf{J}_{M}$ is the $M\times M$ matrix of ones. When the matrix given in Eq.~(\ref{eq:piid}) is applied to the vector $\partial_{\varphi}\mean{\hat{\textbf{o}}}_{\varphi}$, the contribution from $\mathbf{J}_{M}$ vanishes because $\sum_{k}\mean{\hat{o}_{k}}_{\varphi}=\sum_{k}P_{k}=1$, and thus, Eq.~(\ref{eq:picon}) holds. 
The corresponding generalized eigenvalue can be written as
\begin{equation}
\lambda^{-1}=\frac{\partial\mean{\hat{\textbf{o}}}^{T}_{\varphi}}{\partial\varphi}\mathbf{\Gamma}^{+}\frac{\partial\mean{\hat{\textbf{o}}}_{\varphi}}{\partial\varphi}=\sum^{M}_{k=1}\frac{1}{P_{k}}\left( \frac{\partial P_{k}}{\partial\varphi} \right)^{2}=F,
\end{equation}
which corresponds to the Fisher information under coarse-grained measurement. Hence, the optimal weight defined in Eq.~(\ref{eq:optw}) yields the minimum phase variance that saturates the Cramér–Rao bound.

\section{Optimal weight values}
\label{Appendix:B}

The explicit values of the optimal weight vectors in Eq.~(\ref{eq:optw}) for equal-sized and optimal-sized binning configurations are shown in Tables~\ref{tab:weq} and \ref{tab:wopt}.

\renewcommand{\arraystretch}{1}
\begin{table*}[htbp]
\centering
\adjustbox{max width=\textwidth}{
\begin{tabular}{c|c c c c c c c c c c}
~~~$M$~~~ & $w_1$ & $w_2$ & $w_3$ & $w_4$ & $w_5$ & $w_6$ & $w_7$ & $w_8$ & $w_9$ & $w_{10}$ \\
\hline
2  & \(0.707\) & \(-0.707\) \\
3  & \(0.707\) & \(0\) & \(-0.707\) \\
4  & \(0.676\) & \(0.206\) & \(-0.206\) & \(-0.676\) \\
5  & \(0.637\) & \(0.307\) & \(0\) & \(-0.307\) & \(-0.637\) \\
6  & \(0.601\) & \(0.354\) & \(0.116\) & \(-0.116\) & \(-0.354\) & \(-0.601\) \\
7  & \(0.569\) & \(0.376\) & \(0.186\) & \(0\) & \(-0.186\) & \(-0.376\) & \(-0.569\) \\
8  & \(0.542\) & \(0.385\) & \(0.230\) & \(0.076\) & \(-0.076\) & \(-0.230\) & \(-0.385\) & \(-0.542\) \\
9  & \(0.517\) & \(0.387\) & \(0.257\) & \(0.128\) & \(0\) & \(-0.128\) & \(-0.257\) & \(-0.387\) & \(-0.517\) \\
10 & \(0.496\) & \(0.385\) & \(0.275\) & \(0.165\) & \(0.055\) & \(-0.055\) & \(-0.165\) & \(-0.275\) & \(-0.385\) & \(-0.496\) \\
\end{tabular}
}
\caption{Weight vector $\mathbf{w}$ for various bin numbers $M$ under equal-sized binning.}
\label{tab:weq}
\end{table*}

\begin{table*}[htbp]
\centering
\adjustbox{max width=\linewidth}{
\begin{tabular}{c|c c c c c c c c c c}
~~~$M$~~~ & $w_1$ & $w_2$ & $w_3$ & $w_4$ & $w_5$ & $w_6$ & $w_7$ & $w_8$ & $w_9$ & $w_{10}$ \\
\hline
2  & \(0.707\) & \(-0.707\) \\
3  & \(0.707\) & \(0\) & \(-0.707\) \\
4  & \(0.677\) & \(0.203\) & \(-0.203\) & \(-0.677\) \\
5  & \(0.646\) & \(0.287\) & \(0\) & \(-0.287\) & \(-0.646\) \\
6  & \(0.618\) & \(0.327\) & \(0.104\) & \(-0.104\) & \(-0.327\) & \(-0.618\) \\
7  & \(0.594\) & \(0.347\) & \(0.164\) & \(0\) & \(-0.164\) & \(-0.347\) & \(-0.594\) \\
8  & \(0.572\) & \(0.358\) & \(0.201\) & \(0.065\) & \(-0.065\) & \(-0.201\) & \(-0.358\) & \(-0.572\) \\
9  & \(0.553\) & \(0.362\) & \(0.226\) & \(0.109\) & \(0\) & \(-0.109\) & \(-0.226\) & \(-0.362\) & \(-0.553\) \\
10 & \(0.536\) & \(0.364\) & \(0.242\) & \(0.140\) & \(0.046\) & \(-0.046\) & \(-0.140\) & \(-0.242\) & \(-0.364\) & \(-0.536\) \\
\end{tabular}
}
\caption{Weight vector $\mathbf{w}$ for various bin numbers $M$ under optimal-sized binning.}
\label{tab:wopt}
\end{table*}

\end{document}